\documentclass[epj]{svjour}
\usepackage{color}
\usepackage{graphicx}
\usepackage{graphics}
\usepackage{subfigure}
\usepackage{amssymb,amsmath}
\usepackage{gensymb}
\usepackage{multicol, blindtext, graphicx}
\begin{document}
\title{Electronic properties of single and double napped carbon nanocones }
\author{F.A. Gomes\inst{1}, V.B. Bezerra\inst{1}, J.R.F Lima\inst{2} \and F. Moraes\inst{1,2} 
}                     
%
%
\institute{Departamento de F\'{\i}sica, CCEN,  Universidade Federal 
da Para\'{\i}ba, Caixa Postal 5008,  58051-900, Jo\~ao Pessoa, PB, Brazil \and Departamento de F\'{\i}sica, Universidade Federal Rural de Pernambuco, 52171--900, Recife, PE, Brazil}
\date{Received: date / Revised version: date}
\authorrunning{F. A. Gomes \it{et al.}}
\abstract{
In this paper we study the electronic properties of carbon nanocones with one and two nappes, with pentagonal and heptagonal defects in their lattices. We use the continuum model, which is based on a Dirac-like Hamiltonian with the topological defects described by localized non-Abelian gauge field fluxes. We develop a geometrical approach that can describe the two nappes of the double cone surface simultaneously, by extending the radial coordinate to the complete set of real numbers.  We show that, for some combinations of different nanocones, forming the double conical surface, the local density of states near the apex of the cone does not vanish at the Fermi energy and presents a strong dependence on the angular momentum. We also obtain the energy spectrum for finite-sized nanocones and verify that it depends on the choice of topological defect on the surface, which suggests that a double nanocone can be used to control the electronic transport in carbon-based electronic devices.}
%
%
\maketitle
%



\section{Introduction}
\label{intro}

Carbon nanomaterials have attracted a great deal of attention in the last years due to their unusual physical properties and the wide range of potential applications \cite{Tagmatarchis}. The boom in the study of these materials occurred after the discovery of fullerenes in 1985 \cite{KROTO} and carbon nanotubes in 1991 \cite{Iijima}. Since then, various carbon nanomaterials have been obtained, such as nanocones \cite{GE}, graphene \cite{Novoselov}, nanoscrolls \cite{doi:10.1021/nl900677y}, onions \cite{Ugarte} and nanotori \cite{Liu}. In particular, carbon nanocones were first observed experimentally in 1992 as endcaps of carbon nanotubes \cite{Iijima2,Iijima3}, and in 1994 as free-standing structures \cite{Maohui,GE}. Carbon nanocones with cone angles of 19$^{\circ}$, 39$^{\circ}$, 60$^{\circ}$, 85$^{\circ}$ and 113$^{\circ}$ have already been observed \cite{Krishnan}. The carbon nanocone is a result of the introduction of a topological defect called disclination in a planar graphene sheet, resulting from the substitution of a hexagon by either a pentagon or a heptagon in the graphene lattice. 

The study of the electronic properties of carbon  nanocones  revealed an enhancement in the local density of electronic states (LDoS) in the vicinity of the cone apex, which was obtained by first principles  \cite{PhysRevLett.78.2811,PhysRevB.61.8496,PhysRevLett.86.5970}, tight-binding \cite{PhysRevB.49.7697,PhysRevB.52.6015,PhysRevB.64.195419} and continuum model \cite{lammert2000topological,lammert2004graphene} calculations. This enhancement can be used in applications of carbon nanocones in field emission \cite{0957} and scanning probes \cite{PhysRevLett.86.5970}, for instance. A carbon nanocone-based electronic rectification device was also proposed \cite{doi:10.1063/1.3684276}, and it was demonstrated that carbon nanocones are excellent thermal rectifiers \cite{doi:10.1063/1.3049603}, which makes them a promising practical phononic device. Carbon nanocones have also been suggested as  cheaper and more easily produced alternative to carbon nanotubes for applications as gas storage devices \cite{doi:10.1021/jp2069094} and as capsules for drug delivery \cite{doi:10.1021/nn800395t}. 

Another kind of conical structure that has been attracting attention in the last few years is the double cone, where  two cones are connected by their apex. Theoretical studies indicate interesting properties like a high dependence of the electronic states on the angular momentum of the electrons, both for the classical and for non-relativistic quantum dynamics problem \cite{kowalski2013dynamics}. In Ref. \cite{1751-8121-50-6-065302}, it is  presented the relativistic quantum problem of a charged particle restricted to a double cone with and without a magnetic field. Even though these studies have shown interesting properties for the double cone structure, the literature for carbon double nanocones is still very scarce. In a computational simulation of carbon double nanocones it was indicated that its experimental realization is possible, since its formation energy is slightly lower than the energy of a single cone  \cite{lopes2015theoretical}. These results motivated us to explore other possible conical defects in carbon structures and investigate their correspondent electronic properties.

In this paper, we study the electronic properties of carbon nanocones for single and double conical surfaces. We develop and use a continuum model to describe the double conical surfaces, which is an extension of the continuum model developed in Ref. \cite{lammert2004graphene} for nanocones with only one nappe. Even though this model is limited to low-energy electronic states near the Fermi energy, it is very useful in long-distance physics, in which situation it is difficult to deal with first-principles methods due to the great number of atoms. The approach used here allows us to explore many possible combinations of defects in the nanocone with one or two nappes. We focus our attention in pentagonal and heptagonal defects and verified that the connection of two carbon nanocones by their vertexes, creating a double conical surface, brings up new electronic properties that could be used in future applications. 

The paper is organized as follows. In Section \ref{sec:2} we build the continuum model for a double conical surface,  based on the effective Dirac equation for a graphene sheet, with localized gauge fluxes which describe the defects needed to create the nanocone. We solve the effective Dirac equation in Section \ref{sec:3}, and investigate the LDoS and the energy spectrum in Section \ref{sec:4}, showing that it has a direct dependence on the angular momentum and the kind of defects that are present on the surface. The paper is summarized and concluded in Section 5.



\section{The continuum model for a double carbon nanocone}
\label{sec:2}

In this section we will develop a continuum model to describe a double carbon nanocone. The starting point is the effective Dirac equation for low-energy states in graphene. In this model, the lattice of the structure disappears and the pentagons and heptagons in the structure are included in the Dirac equation as localized fictitious gauge fluxes. This model was first proposed in Ref. \cite{PhysRevLett.69.172} to investigate the electronic structure of fullerenes, and has been widely used to describe carbon-based nanostructures \cite{lammert2000topological,lima,jonasjap,Cunha2015}. 

\begin{figure}[hpt]
\centering
 \subfigure[]{ 
\includegraphics[width=3.9cm,height=3.9cm]{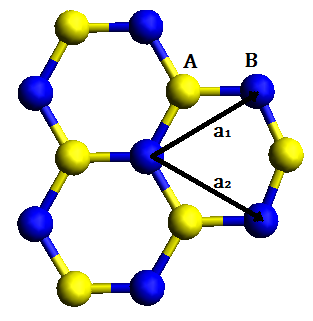}
}
\subfigure[]{ 
\includegraphics[width=3.5cm,height=5cm]{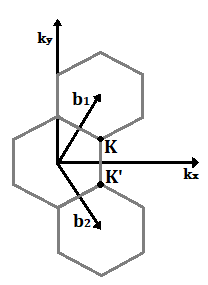}
}
\caption{(Color online) In (a) we see the hexagonal structure of graphene built by two triangular sublattices $A$ and $B$. The lattice vectors are $\vec{a_1}$ and $\vec{a_2}$. (b) is the correspondent Brillouin zone with reciprocal-lattice vectors   $\vec{b_1}$ and $\vec{b_2}$ in reciprocal space. The Dirac points are localized at the points $K$ and $K^{'}$. }
\label{figure2}
\end{figure}

The low-energy electrons in graphene are modeled as massless Dirac fermions obeying the effective equation

\begin{equation}
- i\hbar v_f \sigma^\mu \partial_\mu \Psi = E\Psi ,
\label{masslessdirac}
\end{equation}    
where $v_f$ is the Fermi velocity and the pseudospin operator $\sigma^\mu$ are the usual Pauli matrices, which act on the spinor $\Psi = (\Psi_A, \Psi_B)^T$.  These components are labeled with respect to the sub-lattices, named $A$ and $B$ shown in figure \ref{figure2}. Each component $A$ and $B$ has two subcomponents related to two independent points, $K$ and $K^{\prime}$, in the first Brillouin zone (FBZ), in reciprocal space. These are the points where the band crosses the Fermi level. In other words,  the low-energy excitations  are centered around these two points. Corner points of the FBZ at 120$\degree$ from them are equivalent.  We notice an important covariance when  the coordinate frame ($\hat{e}_\mu$, defined for $K$) is rotated counterclockwise by $\theta$. Under this rotation, the wavefunction is  acted upon by $exp(\frac{i\theta \sigma^3}{2})$, which preserves the Dirac Hamiltonian. We observe the same behavior for $K^{'}$, but the matching frame is rotated by 180$\degree$ from the corresponding one at K. In order to avoid this inconvenience,  the $K^{'}$ frame is rotated in order  to coincide with the $K$ frame. The dispersion relation is the same in these two points, and this degree of freedom is called K-spin \cite{neto2009electronic}. The Pauli matrices acting on the pseudospin and on the K-spin are labeled by $\sigma^\mu$   and $\tau^\mu$, respectively. 

In this paper, we focus our attention on graphene conical surfaces with one and two nappes. The  dynamics of quantum particles on this kind of surface has been studied in recent years due to its  high dependence on the angular momentum \cite{kowalski2013dynamics} and also to the possibility of experimental realization \cite{lopes2015theoretical}. 

While a single cone can be simply described in spherical coordinates $(r,\theta,\phi)$ by $\theta=const.$, the double cone requires two values for $\theta$. In order to work with this surface, we  modify the spherical coordinate system to keep a constant $\theta$ as the double cone equation. \cite{1751-8121-50-6-065302}. We do this by extending the radial coordinate to the whole set of the real numbers. The differential distance vector in this case is given by

\begin{eqnarray}
d\vec{r} = dl \ \hat{e}_{l} + \left| l \right| \sin \theta d\phi \ \hat{e}_{\phi},
\label{lineelement}
\end{eqnarray}
where $l \in \mathbb{R}$ is the new radial coordinate.
The square of the line element is then

\begin{equation}
ds^2 = d\vec{r} . d\vec{r} = dl^2 + \left| l \right|^2\alpha^2d\phi^2 ,
\label{metric}
\end{equation}
where $\alpha = \sin\theta$. 

\begin{figure}[hpt]
\centering
	\includegraphics[width=6cm,height=5cm]{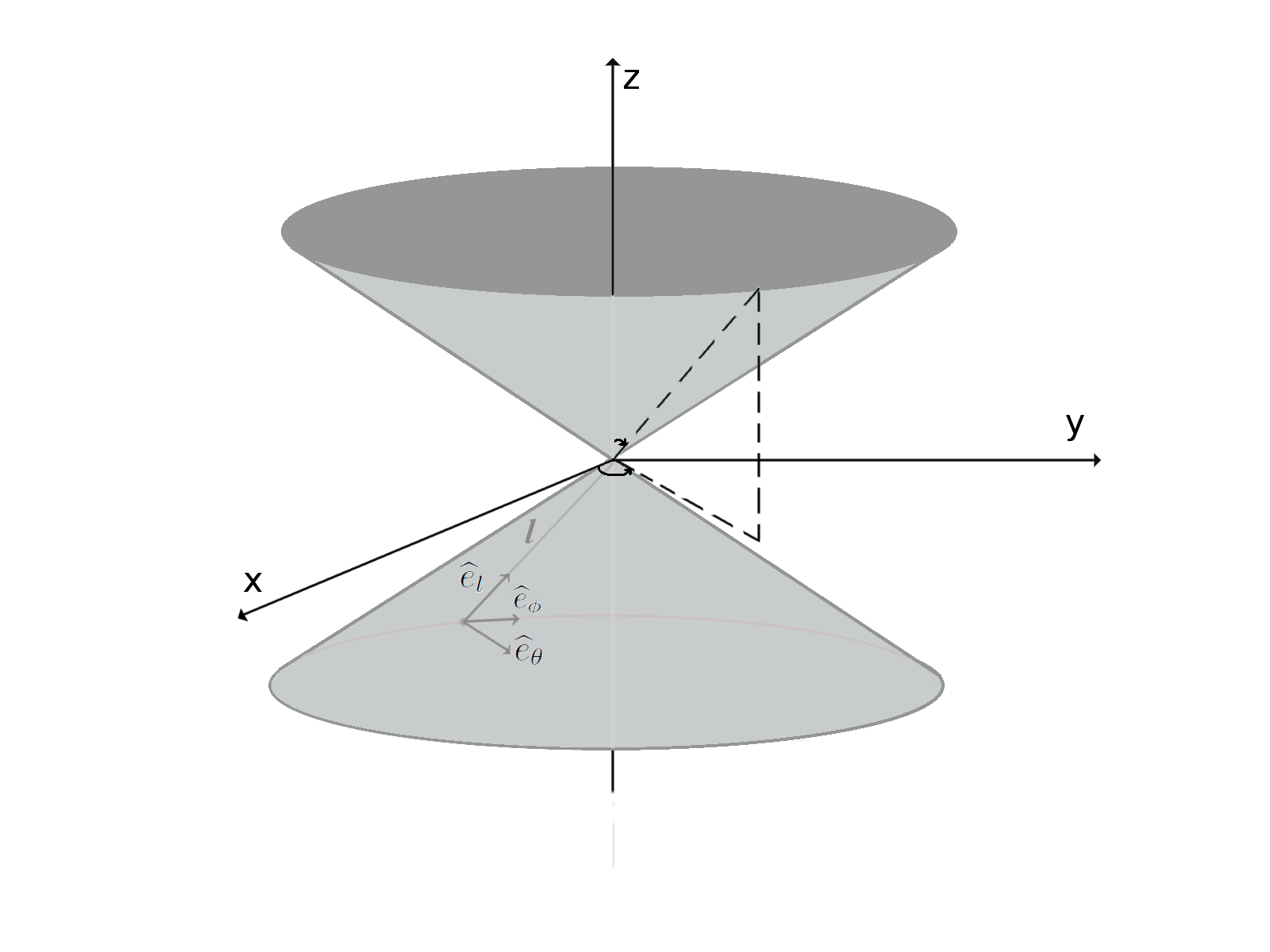}
\caption{Coordinate system for a double cone surface.}
\label{figure2.1}
\end{figure}

The tetrad formalism is specially useful \cite{birrell1984quantum,misner1973gravitation} to deal with the relation (\ref{metric}) and the Dirac equation (\ref{masslessdirac}).  Using the tetrad formalism, we introduce at each point a set of locally inertial coordinates and a set of orthonormal vectors, $e_\mu^a$, fixing the transformation between the local (Latin indexes) and the general (Greek indexes) coordinates. These vectors are found by the relation $g_{\mu\nu}(x)=e^a_\mu(x) e^b_\nu(x) \eta_{ab}$, where $g_{\mu\nu} = diag(1,\left| l \right|^2\alpha^2) $ and $\eta_{ab} = diag(1,1)$. In our problem they are given by

\begin{equation}
e^\mu _a = \left(\begin{array}{cc}
1 & 0  \\
0 & \frac{1}{\alpha \left| l \right| } \end{array} \right) .
\label{gammacurved2}
\end{equation}

The $\sigma^\mu$ matrices in (\ref{masslessdirac}) are related with the flat space matrices $\sigma^a$ by $ \sigma^\mu = e^\mu _a \sigma^a$. Therefore

 \begin{eqnarray}
\sigma^l & = & \sigma^1 , \label{gamma} \\
\sigma^\phi & = & \frac{\sigma^2}{\alpha  \left| l \right| } \nonumber .
\end{eqnarray}

In order to give the shape of a cone to the flat hexagon-tiled surface of graphene one can substitute one of its hexagons for a pentagon. This leads to the appearance of a Aharonov-Bohm-like phase in the wavefunction of quantum particles moving around it \cite{furtado2008geometric}. This effect can  be incorporated into the Hamiltonian of the continuum model by the addition of two fictitious non-Abelian gauge fields piercing the surface at the location of the pentagon. One of them  promotes the frame rotation covariance and the other one, the $K$ spin rotation invariance to local symmetries. The procedure is similar to the electromagnetic gauge potential transformation used to compensate a change of wave function phase in the Aharonov-Bohm effect \cite{aharonov1961further}. In the case described in this article a spinorial connection is added to the covariant derivative in order to accomplish the frame rotation covariance \cite{1751-8121-50-6-065302}. Thus, we introduce the spin connection, $\Omega_\mu = - \frac{1}{8} w_{\mu, a, b} [\sigma^a, \sigma^b] $, in the covariant derivative $ \nabla_\mu = \partial_\mu + \Omega_\mu $, where $w_{\mu, a, b}$ is the spinorial connection.  This gauge field can be explicitly expressed as

\begin{equation}
\Omega _\phi = -\frac{i}{2}\alpha \sigma^3 .
\label{spinorialconn}
\end{equation}
where $\sigma^3$ is the third Pauli matrix acting on the components of the spinor related to the sublattices A and B. To promote  $K$ spin rotation invariance, we introduce the also non-Abelian gauge field, for one and two defects, given by \cite{lammert2004graphene}

\begin{equation}
a_{\phi} = - \frac{3}{2} \frac{\left(\alpha -1 \right)}{\alpha  \left| l \right| } \tau^2 
\label{gauge1}, \quad \text{for one defect,}
\end{equation}
and
\begin{eqnarray}
a_{\phi} = \frac{-3\left( \alpha - 1 \right)}{2\alpha  \left| l \right|} \left[1 - \frac{2}{3} \left(n-m\right)\right] \tau^3, \label{gauge2} \\ \nonumber \text{for two defects.}
\end{eqnarray}

We  see in the relation (\ref{gauge2}), for two defects, a dependence with $n$ and $m$, which are the parametric coordinates (n,m) of the defects' centers in the graphene lattice, as defined in \cite{lammert2004graphene}. 

The continuum model developed in this section allows us to investigate the electronic properties of a variety of graphene conical surfaces. In the next section, we use this approach to solve the free particle problem in this geometric approach.



\section{Free particle problem}
\label{sec:3}

In this section, we solve the effective Dirac equation developed in the last section for the case of a free particle constrained to a carbon conical surface. With this solution, we will be able to evaluate the influence of the surface's geometry on its electronic properties. We start by rewriting the effective Dirac equation \eqref{masslessdirac} including now the non-Abelian gauge fields, 

\begin{eqnarray}
\left[ \sigma^i p_i - i\hbar \sigma^i \Omega_i - \hbar \sigma^i a_i  - \frac{E}{v_f} \right] \psi = 0.
\label{effDirac0}
\end{eqnarray}
Choosing to work with the quadratic form of this equation, we apply on it the operator

\begin{eqnarray}
\left[ \sigma^j p_j - i\hbar \sigma^j \Omega_j - \hbar \sigma^j a_j + \frac{E}{v_f} \right] 
\end{eqnarray}
\label{operator}
and after some calculations, we get 

\begin{eqnarray}
p_i &p_i& \psi - 2 i \hbar \Omega_i p_i \psi - 2 \hbar a_{i} p_i \psi - \hbar^2 \Omega_i  \Omega_i \psi 
 \nonumber \\
+ &2& i \hbar^2 \Omega_i a_{i}  \psi + \hbar^2 a_{i}a_{ i}  \psi - \frac{E^2}{v_f^2} = 0.
\label{effDirac1} 
\end{eqnarray}
Using the $\sigma^\mu$ matrices (\ref{gamma}), the spinorial connection (\ref{spinorialconn}) and considering that $a_i$ only has component in $\phi$, we get

\begin{eqnarray}
&-&  \frac{\partial^2}{\partial l^2} \psi - \frac{1}{ l } \frac{\partial}{\partial l} \psi - \frac{1}{l^2 \alpha^2} \frac{\partial^2}{\partial \phi^2} \psi + i \frac{\sigma^3}{\alpha l^2} \frac{\partial}{\partial \phi} \psi  + \frac{2i a_{\phi}}{\alpha \left|l\right|} \frac{\partial}{\partial \phi} \psi
  \nonumber \\
&-& \frac{1}{4 l^2} \psi + \frac{ a_{\phi} \sigma^3}{\left| l \right|} \psi + {a_{\phi}}^2 \psi  - k^2 \psi = 0,
\label{effDirac2}
\end{eqnarray}
where we defined 

\begin{equation}
k^2 = \left(\frac{E}{\hbar v_f}\right)^2 .
\label{k}
\end{equation}

The solutions of the above equation  have the form 
\begin{equation}
\psi(l,\phi) = e^{ij\phi}\psi_j(l),
\end{equation}
where $\psi_j(l)$ is a function of  the coordinate $l$ only and the angular momentum quantum number $j$ is an integer plus one-half. 

We diagonalize Eq. (\ref{effDirac2})  taking into account that the only operators acting on the $K$ spin are $\tau^2$ and $\tau^3$ from the gauge fields (\ref{gauge1}) and (\ref{gauge2}), respectively.  Similarly, the matrix $\sigma^3$ that acts on the pseudospin is
replaced by its eigenvalue $\sigma = \pm 1$, where the plus (minus) sign is related to the sub-lattice A (B). Equation \eqref{effDirac2} then becomes

\begin{eqnarray}
&-& \frac{\partial^2}{\partial l^2} \psi_j - \frac{1}{ l } \frac{\partial}{\partial l} \psi_j + \frac{j^2}{l^2 \alpha^2} \psi_j - \frac{j\sigma}{\alpha l^2} \psi_j - \frac{2j \tau a^{'}_{\phi}}{\alpha l^2} \psi_j
  \nonumber  \\
&+& \frac{1}{4l^2} \psi_j + \frac{ a^{'}_{\phi} \sigma \tau}{l^2} \psi_j 
 + \frac{{a^{'}_{\phi}}^2}{l^2} \psi_j - k^2 \psi_j = 0,
\label{effDirac3}
\end{eqnarray}
where $\tau = \pm 1$, 

\begin{eqnarray}
a^{'}_{\phi} = - \frac{3}{2} \frac{\left(\alpha -1 \right)}{\alpha }, \quad \text{for one defect,}
\label{gauge11}
\end{eqnarray}
and
\begin{eqnarray}
a^{'}_{\phi} = - \frac{3}{2}\frac{\left( \alpha - 1 \right)}{\alpha } \left[1 - \frac{2}{3} \left(n-m\right)\right], \label{gauge21} \\ \nonumber
\text{for two defect,}
\end{eqnarray}
in order to make explicit the radial coordinate ($ l $) dependence due to the  gauge fields. We  reorganize the terms in Eq. (\ref{effDirac3}) and rewrite it in the form

\begin{eqnarray}
\frac{\partial^2}{\partial l^2} \psi_j + \frac{1}{ l } \frac{\partial}{\partial l} \psi_j - \frac{\nu^2}{l^2} \psi_j + k^2 \psi_j = 0,
\label{effDirac4}
\end{eqnarray}
where

\begin{equation}
\nu^2 = \left( \frac{j}{\alpha} - \tau a^{'}_{\phi} - \frac{\sigma}{2} \right)^2.
\label{Nu}
\end{equation}

Equation (\ref{effDirac4}) is clearly a Bessel equation, whose  solution is a combination of Bessel functions of first and second kinds, respectively. Since we are including the origin ($l =0$), we do not consider the Bessel function of second kind because it diverges there. Thus the solution is written as

\begin{equation}
\psi_j(l) = C J_\nu(kl),
\label{psiA}
\end{equation}
where $C$ is a normalization constant and $J_{\nu} (kl)$ is the Bessel function of  first kind.  It is important to notice that the results obtained here reproduce the known results for  flat graphene and for the singly napped carbon nanocone \cite{lammert2004graphene,lammert2000topological,bueno2012landau}. If we consider $\alpha = 1$, the gauge fields (\ref{gauge1}) and (\ref{gauge2}) vanish giving rise to a Bessel function of  index  $ j - \frac{1}{2}$, as expected for the flat case.

At this point, we  notice that the system of coordinates used here has the advantage of allowing us to deal with two single cones, joined by their apexes,  in a unified form. Since we have two single nanocones in the same structure, we expect to reproduce the known results for a single carbon nanocone, but in a  domain twice as large. This is exactly what we obtain in this section. The solution is in agreement with the known results for a single cone, but now it is dependent on the extended radius $l$.

This unified system of coordinates brings us the possibility of putting together different conical structures and, in this way,  combine their properties. In the next section we will explore the electronic properties of some of these combinations.

\section{Electronic Properties}
\label{sec:4}

\begin{figure}[hpt]
\centering
\subfigure[]{ \includegraphics[width=4cm,height=3cm]{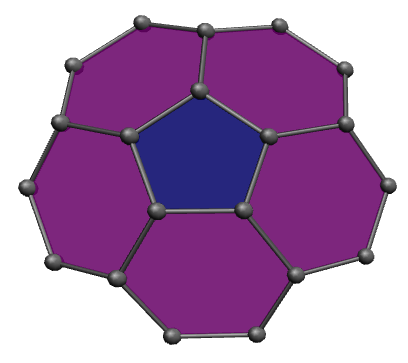}}
\subfigure[]{ \includegraphics[width=4cm,height=3cm]{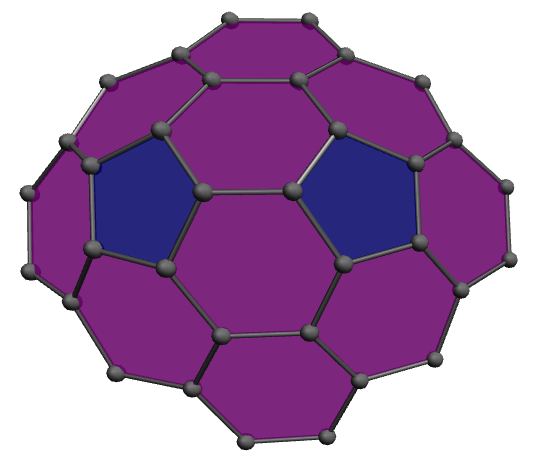}}
\subfigure[]{ \includegraphics[width=4cm,height=3cm]{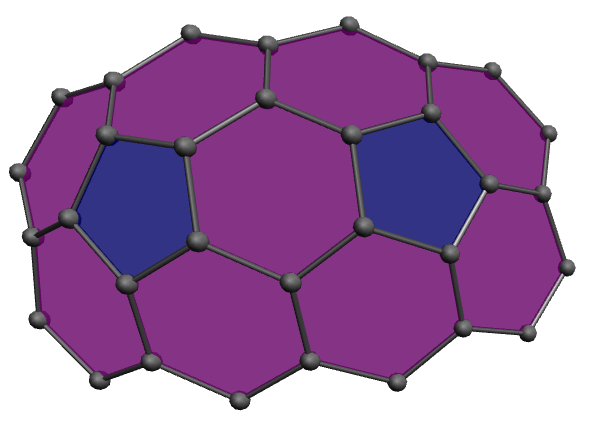}}
\subfigure[]{ \includegraphics[width=4cm,height=3cm]{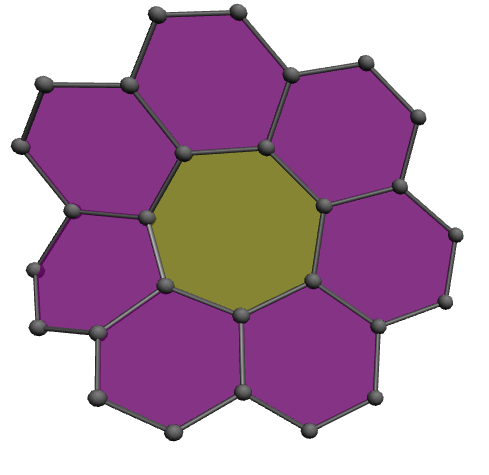}}
\subfigure[]{ \includegraphics[width=4cm,height=3cm]{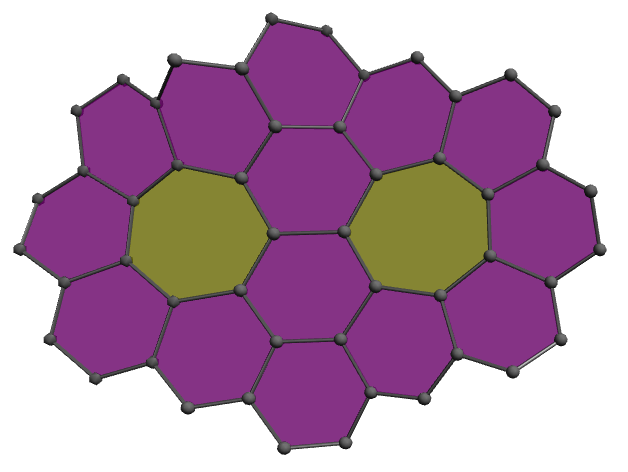}}
\subfigure[]{ \includegraphics[width=4cm,height=3cm]{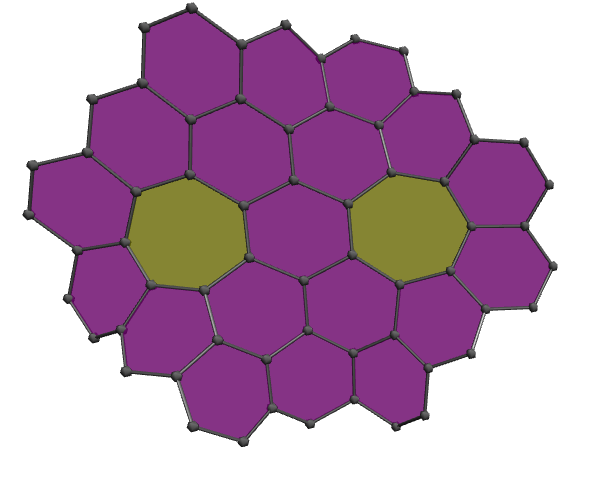}}
\caption{Conical defects on a graphene surface. (a), (b) and (c) are pentagonal defects, wherein (b) has parametric coordinate (1,1) and (c) (2,0). Structures (d), (e) and (f) have heptagonal defects,  wherein (e) has parametric coordinate (1,1) and (f), (2,0).}
\label{figure3}
\end{figure}

The model developed in this article allows us to study a variety of conical surfaces. The first parameter that gives us freedom to model different deformed surfaces is the parameter $\alpha$ which appears in the metric \eqref{metric} and is given by

\begin{eqnarray}
\alpha = 1 + \frac{\lambda}{2\pi} ,
\end{eqnarray}
where $\lambda$ is the angle of the inserted ($\lambda>0$) or removed  ($\lambda<0$) slice from the flat graphene lattice in order to make the cone, in the procedure known as the Volterra process \cite{puntigam1997volterra}.  

The hexagonal symmetry of the graphene sheet requires that $\lambda$  be a multiple of $\frac{\pi}{3}$. Moreover,  square or octagonal defects are rare to occur due to the high deformation they promote and consequently due to their high formation energy \cite{lammert2004graphene}. Therefore, we consider here only pentagonal ($\lambda = - \frac{\pi}{3}$) and heptagonal ($\lambda = \frac{\pi}{3}$) defects. 

\begin{table}[h!]
\centering
\begin{tabular}{|l|r|r|r|r|r|}
\hline
Type of defect & $\alpha$ & $a^{'}_{\phi}$ & $\nu_0$ \\ \hline
1 pentagon & $5/6$ & $3/10$ & $-1/5$  \\ \hline
2 pentagons (1,1) ($n \not\equiv m (mod3)$) & $2/3$ & $1/4$ & $0$  \\ \hline
2 pentagons (2,0) ($n \equiv m(mod3)$) & $2/3$ & $3/4$ & $-1/2$  \\ \hline
1 heptagon & $7/6$ & $- 3/4$ & $-2/7$  \\ \hline
2 heptagons (1,1) ($n \not\equiv m(mod3)$) & $4/3$ & $- 1/8$ & $-1/4$  \\ \hline
2 heptagons (2,0) ($n \equiv m(mod3)$) & $4/3$ & $- 3/8$ & $-1/2$  \\ \hline
\end{tabular}
\caption{Values of $\alpha$ and $a^{'}_{\phi,i}$ for the different structures with pentagonal and heptagonal defects, as illustrated in Fig. \ref{figure3}. Also the values of $\nu_0$, which are the values of $\nu$ that minimize $2\nu+1$ on the given class of the cone.}
\label{table1}
\end{table}

\begin{figure}[hpt]
\centering
\subfigure[]{ \includegraphics[width=3cm,height=3cm]{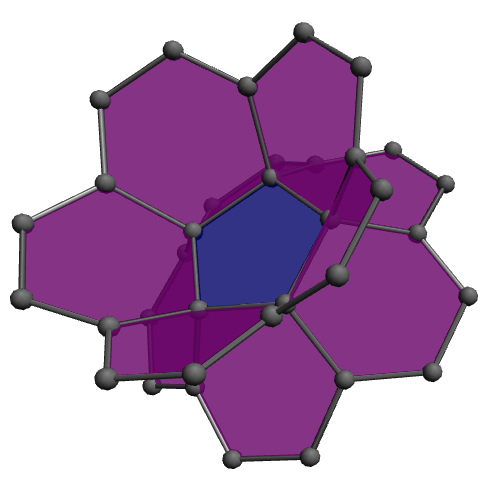}}
\subfigure[]{ \includegraphics[width=4cm,height=3cm]{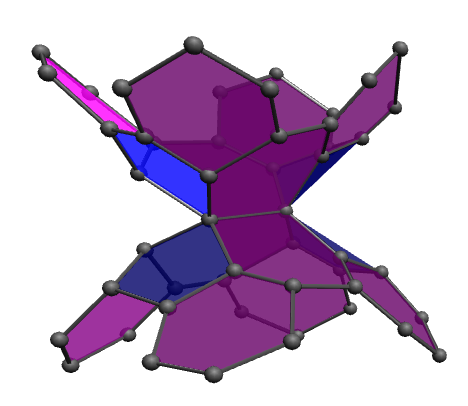}}
\subfigure[]{ \includegraphics[width=4cm,height=3cm]{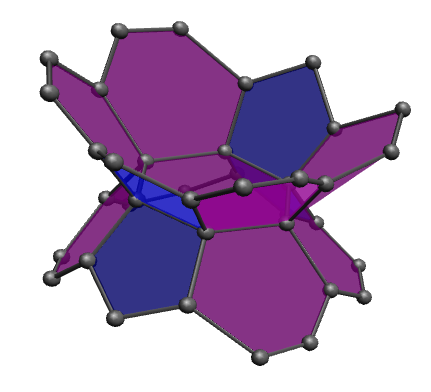}}
\subfigure[]{ \includegraphics[width=4cm,height=3cm]{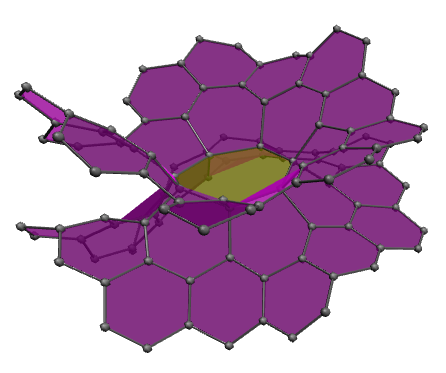}}
\subfigure[]{ \includegraphics[width=4cm,height=3cm]{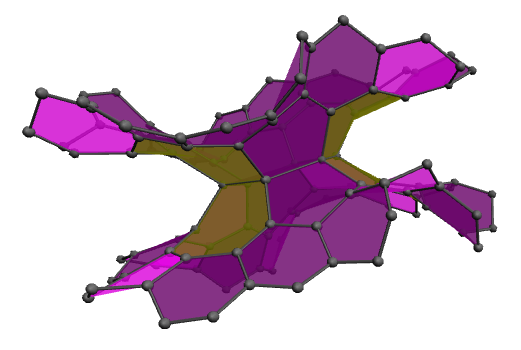}}
\subfigure[]{ \includegraphics[width=4cm,height=3cm]{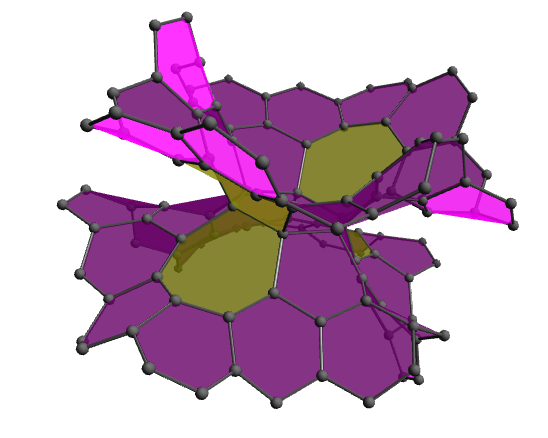}}
\caption{Double conical surfaces built by juxtaposition of the surfaces shown in figure 3.}
\label{figure4}
\end{figure}

Another possibility, is to consider two defects in the same structure. The gauge field (\ref{gauge2}) gives the influence of the defects in terms of their parametric coordinates $(n, m)$. It is known that the effects of topological defects in a graphene surface are dependent on the relative position of the defects. This sensitivity can be classified by a $n-m$ combination rule \cite{lammert2004graphene}. Two classes of defects are sorted depending whether $n-m$ is a multiple of three or not. We choose one example of each class. For the class $n \not \equiv m (mod3)$, we choose the structure (1,1). For the class $n \equiv m (mod3)$, we choose the structure (2,0). So, we consider here six different cones, three with pentagonal and three with heptagonal defects, which are shown in Fig. \ref{figure3}. The values of the parameters $\alpha$ and $a^{'}_{\phi}$ for each cone considered here are given in Table 1. The double conical surfaces that we  consider are combinations of the cones shown in Fig. \ref{figure3}. The combinations with the same cone in each nappe is shown in Fig. \ref{figure4}. 

\subsection{Local density of states}

One important quantity to characterize the electronic and transport properties of a solid is the density of states (DoS). It gives us information about the number of states available to be occupied per interval of energy at each energy level. Thus, it shows us the conducting characteristics   of the material, if it is a metal, insulator or semiconductor. 

Since we are focusing our attention in surfaces with conical defects, which, in the continuum limit, present  singular curvature, we will look more carefully to the local density of states (LDoS) near the apex of the surface. As pointed out in Ref. \cite{lammert2004graphene}, the LDoS in the continuum theory can be written, by dimensional analysis, as

\begin{eqnarray}
\left. \frac{dn}{dE}\right|_l = \frac{f_2 (\left| k \right| l)}{\pi \hbar v_f l} ,
\label{LDoS}
\end{eqnarray}
where $f_2 (\left| k \right| l)$ is a scale function determined from the solution of the effective Dirac equation obtained in the previous section. 

\begin{figure*}[ht]
\centering
\subfigure[]{ \includegraphics[width=6cm,height=6cm]{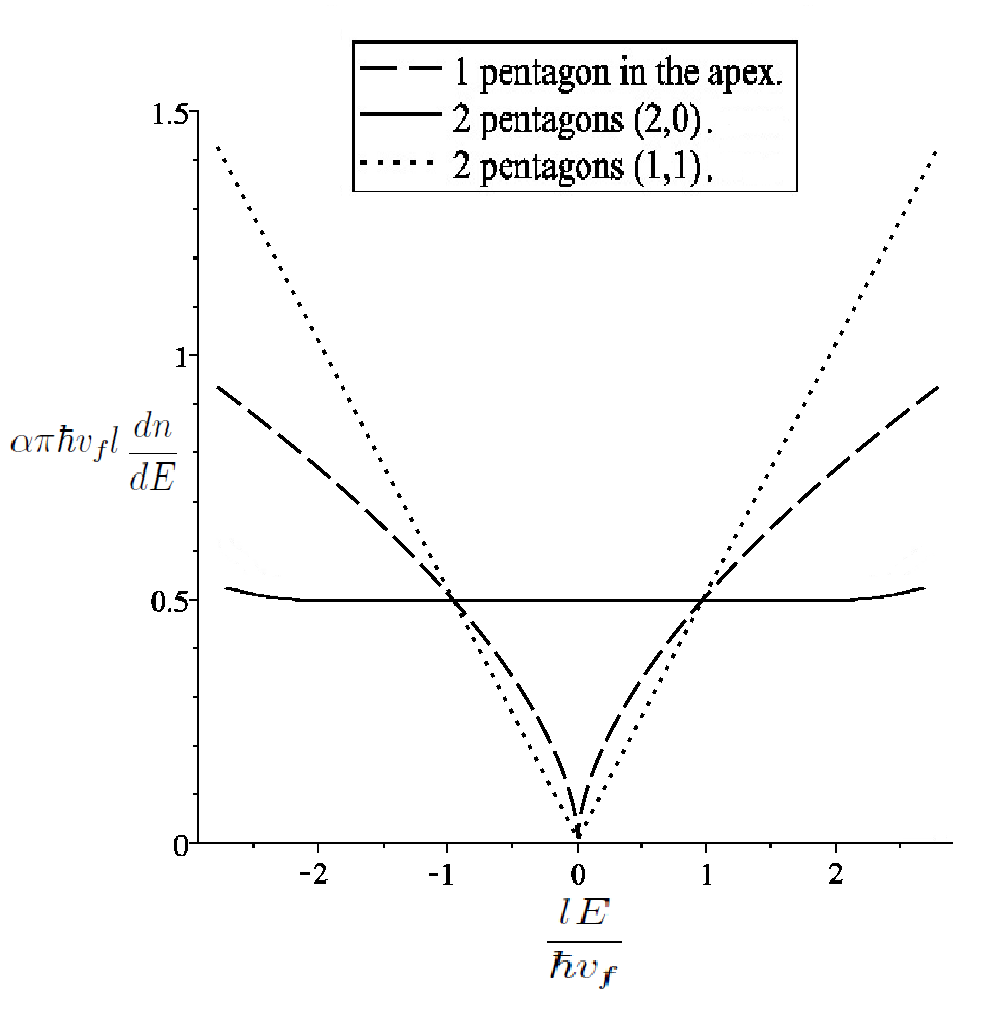}}
\subfigure[]{ \includegraphics[width=5.5cm,height=6cm]{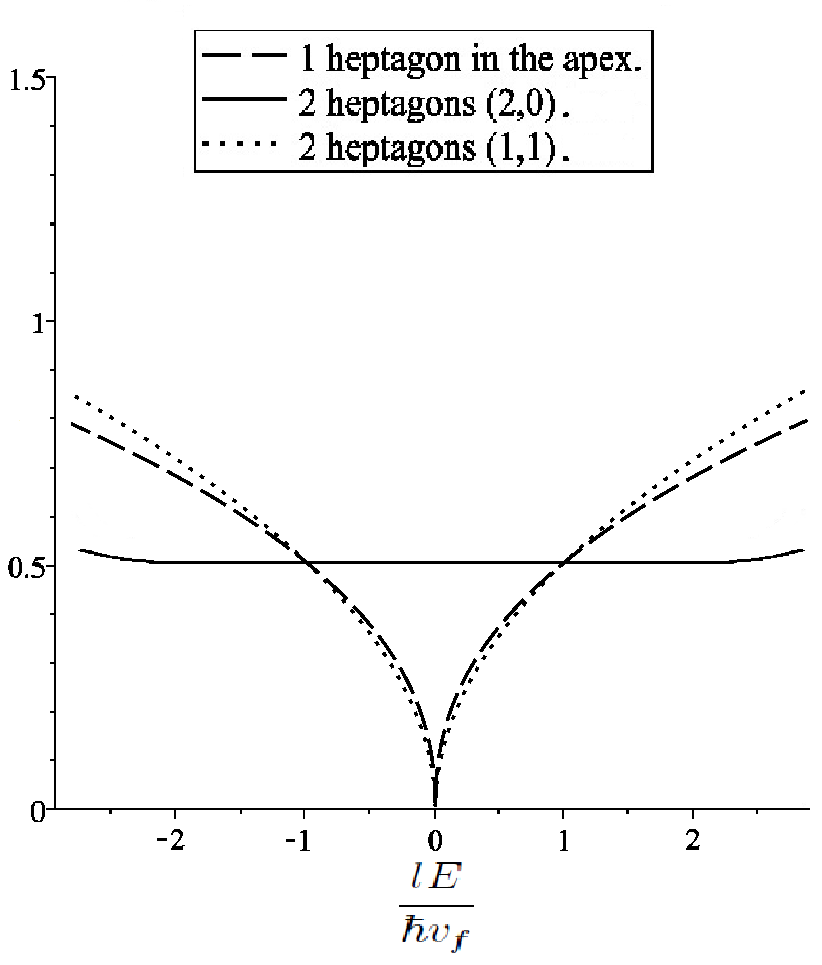}}
\subfigure[]{ \includegraphics[width=5.5cm,height=6cm]{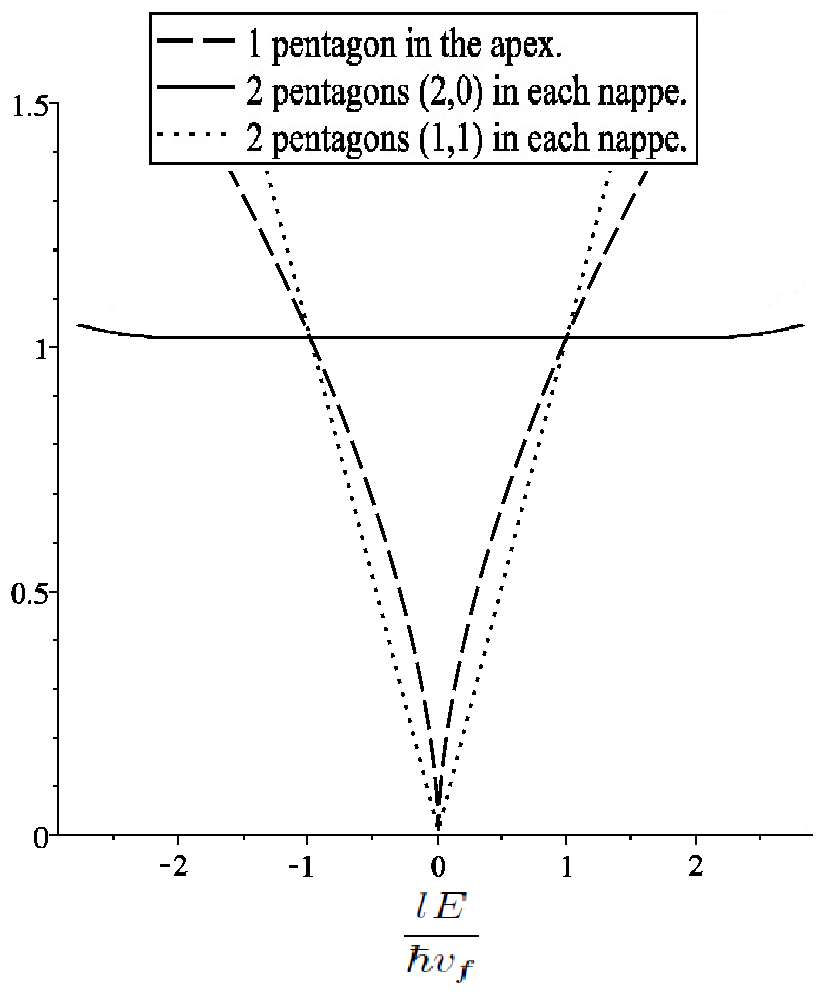}}
\subfigure[]{ \includegraphics[width=6cm,height=6cm]{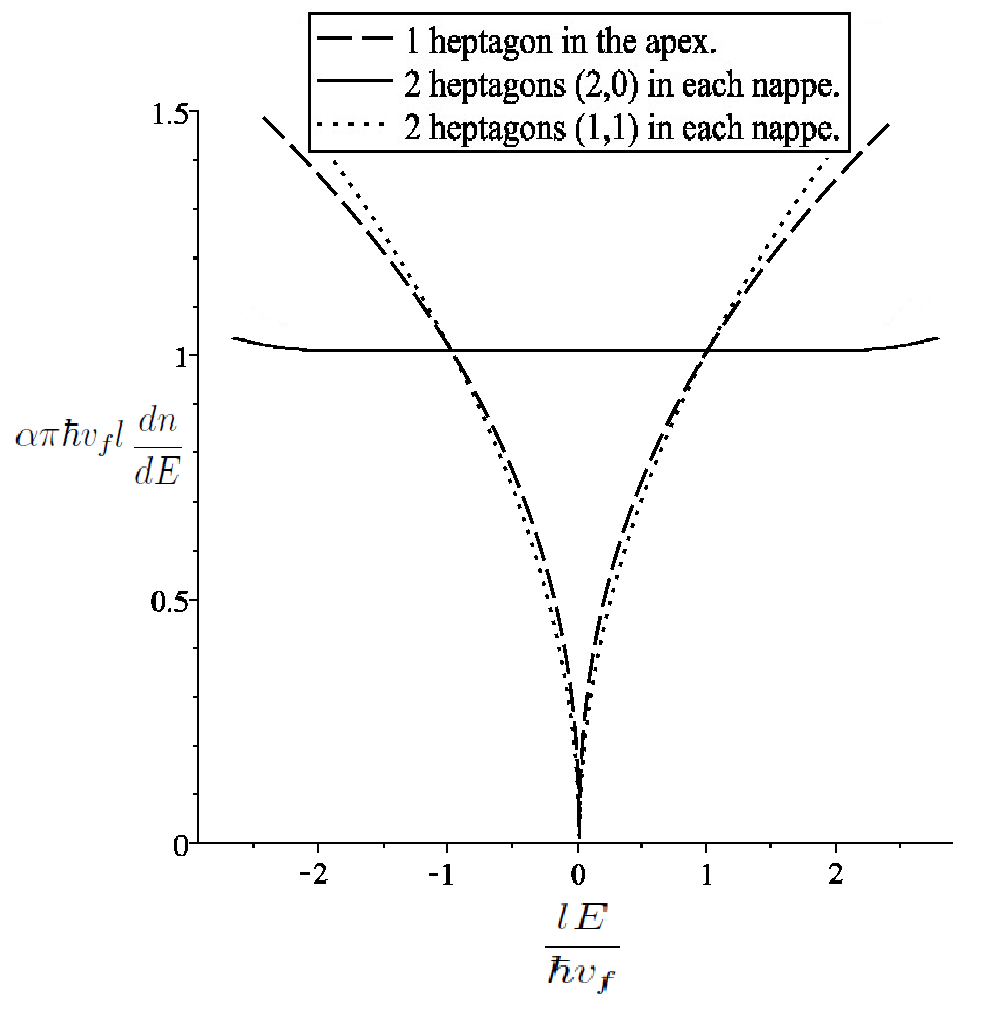}}
\subfigure[]{ \includegraphics[width=5.5cm,height=6cm]{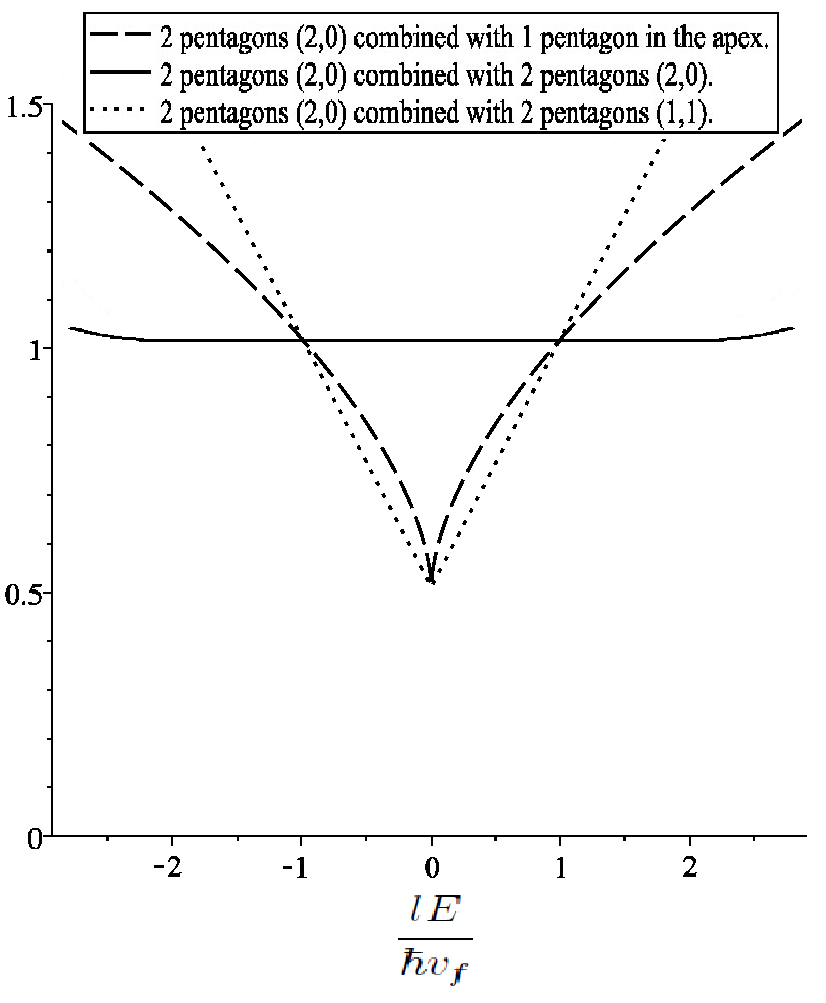}}
\subfigure[]{ \includegraphics[width=5.5cm,height=6cm]{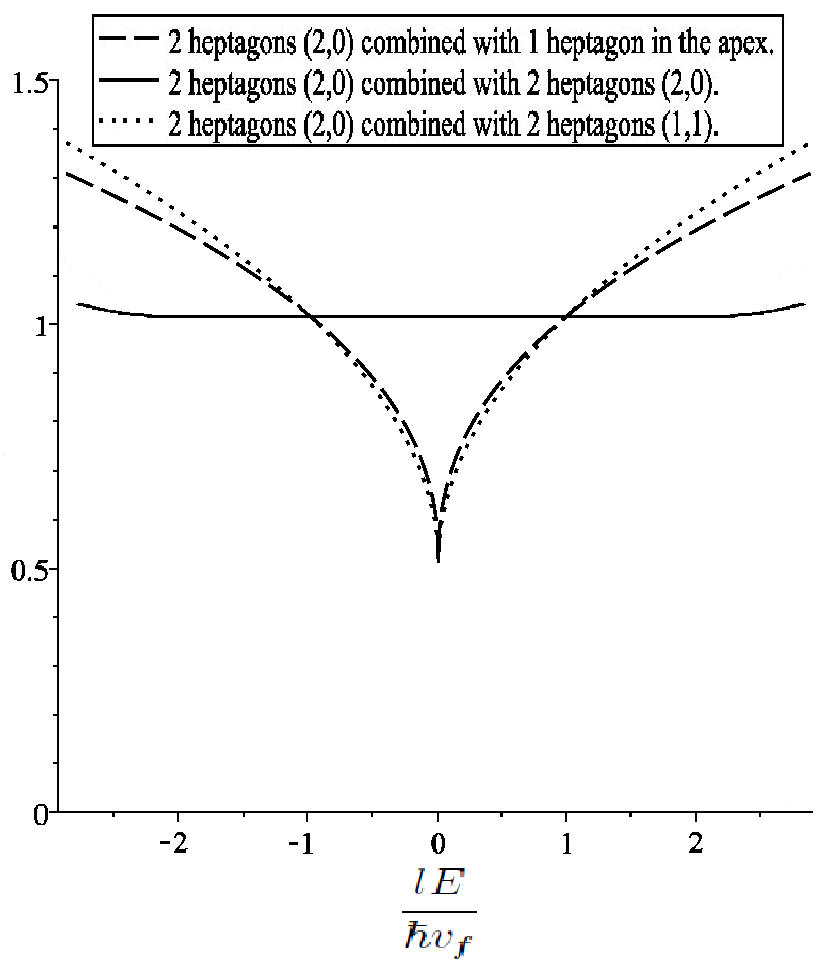}}
\caption{Schematic diagram for the density of states near the apex of the cone, according to equations (\ref{LDoS}) and (\ref{assimp}). We plot the LDoS for cones with pentagonal defects in (a), (c) and (e), where (a) is for single nappe structures, (c) for two identical nappes and (e) for the combination of different nappes with the structure (2,0). We follow the same sequence for heptagonal structures in (b), (d) and (f).}
\label{figure5}
\end{figure*}

In order to obtain an expression for  $f_2 (\left| k \right| l)$, we initially consider a large conical surface with boundary at $\ell=L$ and take $L \rightarrow \infty$.  Taking this condition into account, we impose a boundary condition that ensures that the radial component
of the current density vanishes at the boundary, i.e., $\psi_A(L)\psi_B(L)=0$. This condition leads us to the scale function

\begin{eqnarray}
f_2(\left| k\right| l) = \frac{kl}{2\alpha} \sum_\nu \left| J_\nu (kl) \right|^2 .
\label{ScaleFunction}
\end{eqnarray}
Since we are interested in the LDoS near the apex of the cone, we consider the asymptotic form of the Bessel function as $l \rightarrow 0$ in Eq. (\ref{ScaleFunction}). Then, we obtain

\begin{eqnarray}
f_2(\left| k\right| l) = \frac{kl}{2\alpha} \frac{\left| kl \right|^{2\nu_0}}{2^{2\nu_0} \Gamma (\nu_0+1)^2} .
\label{assimp}
\end{eqnarray}
As k is proportional to the energy E (see Eq. \ref{k}), the LDoS will be proportional to $\left|E\right|^{2\nu_0 +1}$, where $\nu_0$ is the value of $\nu$ that minimizes $2\nu + 1$ on the given cone class. For a given class, the allowed values of $\nu$ (\ref{Nu}) will depend on the parameters $j$, $\tau$ and $\sigma$. The combination of these parameters that minimizes $\nu$, defines $\nu_0$. The values of $\nu_0$ for all cones considered here are in Table \ref{table1}.

The LDoS in terms of the energy is plotted in Fig. \ref{figure5}. We consider structures with pentagonal and heptagonal defects, in single and double carbon nanoconical surfaces, combining different defected structures. It is important to mention that, for all the cones considered here, the values of the angular momentum quantum number $j$ that minimizes $2\nu_0 + 1$ is always $\pm 1/2$, which are the states with zero orbital angular momentum. This is due to the fact that we are looking at the LDoS near the cone's apex, and these states are more concentrated in this region. 
 
The LDoS for cones of a single nappe with pentagonal and heptagonal topological defects are plotted in Fig. \ref{figure5} $(a)$ and $(b)$, respectively. As can be seen, the LDoS shows a direct dependence with the type of defect on the surface. In Fig. \ref{figure5} $(a)$ we reproduce the results obtained in Ref. \cite{lammert2004graphene}, which shows that the cone with (1,1) pentagonal defect has the same linear LDoS near the Fermi energy as  planar graphene, while in the structure (2,0) the LDoS is nearly flat and nonzero at the Fermi energy. As can be seen in Fig. \ref{figure5} $(b)$, for the case of heptagonal topological defects, the flattening of the LDoS near the Fermi energy for the structure (2,0) also occurs, but the linear LDoS of planar graphene is not reproduced by any of the cones. The enhancement of the LDoS observed in the structure (2,0)  occurs because the wave function becomes more strongly concentrated at the apex as $ 2\nu+1$ decreases. The class that presents the minimum of $2\nu + 1$, when we compare $\nu_0$ for all classes, is  $ n \equiv m (mod3)$. We are studying as example of this class, the structure (2,0). In this structure $\alpha = 2/3$ and $a^{'}_{\phi} = 3/4$, this leads to $\nu_0 = -1/2$, and as consequence $2\nu_0 +1 =0$. The same happens for the structure (2,0) with two heptagonal defects, where $\alpha = 4/3$ and $a^{'}_{\phi} = -3/8$. In these cases, the topological defects maintain the conducting characteristic (\textit{i.e.} semimetal) of the structure.

The LDoS of the cones with two nappes are plotted in Figs. \ref{figure5} $(c)$ - $(f)$. In Figs. \ref{figure5} $(c)$ and $(d)$ we consider the cones with the same topological defect in each nappe. The LDoS in this case does not change its behavior, and the result is just the double of the single nappe structure, as expected. A more interesting result comes when the double cone is the combination of two nappes with different defects in the structure, as illustrated in Figs. \ref{figure5} $(e)$ and $(f)$. In this case, the LDoS is the combination of the behavior of the two different nappes. We choose as example, combinations where at least one nappe is a (2,0) structure. As a result, the (2,0) structure leads to a nonzero LDoS  at the Fermi energy, and therefore metallic behavior, in all  cases.

\subsection{Energy spectrum: finite nanocones}

If we consider finite cones with size  $l=L$, the wave function for $l \geq L$ is equal to zero. So, the boundary condition at the edge of the cone is given by
\begin{equation}
\psi_j(L)= C J_{\nu}(kL)=0.
\end{equation}
Then, the argument of the above Bessel function, $kL$, has to be a zero of this function. This gives rise to a discrete energy spectrum given by
\begin{equation}
E_{n,j}=j_{\nu,n}\frac{\hbar v_F}{L}
\end{equation}
where $j_{\nu,n}$ is the $n$'th zero of the Bessel function of order $\nu$.

\begin{figure}[hpt]
\centering
	\includegraphics[width=\linewidth]{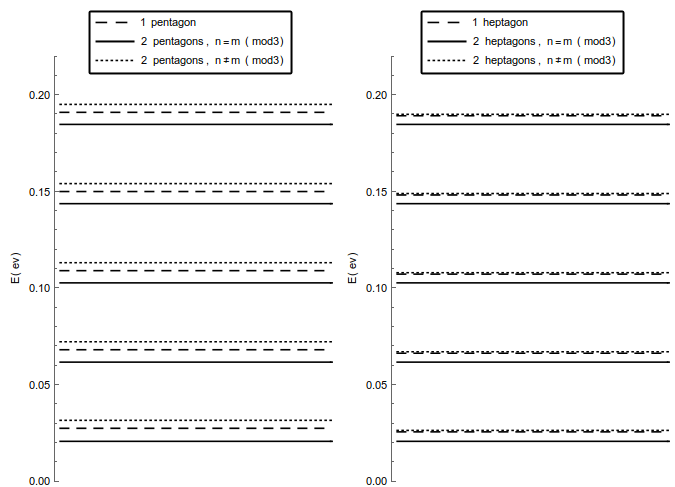}
\caption{The energy spectrum for the pentagonal and heptagonal finite conical surfaces considered here. For all the cases, we consider $L=50$ nm and $\nu=\nu_0$. }
\label{finito}
\end{figure}

In Fig. \ref{finito} we plot the energy spectra for the finite case of the conical surfaces studied in the previous section. We consider  spectra corresponding to the lowest value of $j$, which gives a wave function that is more concentrated in the cone apex. For this case, $\nu=\nu_0$. It is possible to see that the energy spectrum of the cone depends on the topological defects present on the surface, even though  different cones may have the same energy spectrum, as for the case of two pentagons and two heptagons $n=m (mod3)$. We remark that the size of the cone, $L$, can be used to control the interval between subsequent allowed values of energy.

For the case of a double conical surface we  have two values for $L$, one for each cone. Since the energy spectrum depends on $L$ and on the topological defect in the cone, there are many possible combinations of cones with the most varied energy spectra. Thus, one can choose two specific cones in a double conical surface in order to control the electronic transport from one to the other. From the classical point of view, the two cones are connected only by the states with zero angular momentum. In the quantum regime, the lowest angular momentum states are more concentrated in the cones' junction/apex, so they will have a higher probability to move between the cones. In the continuum limit approach that we are using, the apexes of the cones are only a point and we do not have the effects of the connection between the cones. So, further analysis with other approaches, such as tight binding and first principle calculations, are necessary in order to understand the transport properties in the double conical surface.

\section{Concluding remarks}
\label{sec:Conclusions}
 
The study of double carbon nanocone's electronic properties is the purpose of this work. In the continuum model, the nanocones electronic properties are modeled by an effective Dirac equation with the topological defects in the lattice represented by localized gauge fluxes. Each kind of defect is represented by a specific gauge field and multiple defects can be combined in a different net flux. This freedom allows us to explore a variety of defected surfaces, namely, nanocones with one and two pentagons or heptagons in the structure. We use a geometrical approach that makes possible to describe the two cones simultaneously. This approach extends the radial coordinate to the whole set of real numbers.

The LDoS for different combinations of nanocones in the double conical structure is investigated. It is shown that the class of nanocones with two defects and parametric coordinates $ n \equiv m (mod3)$ leads to an enhancement of the apical wave function concentration, resulting in a non-vanishing LDoS at the Fermi energy. We also obtain the energy spectra for finite nanocones, which depends both on their size and on the topological defects on the cone. So, the combination of two nanocones in a double conical surface with suitable topological defects can be used, for instance, to control the electronic transport  from one cone to the other.

The approach developed in this work can be used to explore the electronic properties of other Dirac materials with topological defects. It would be interesting, for future applications of carbon-based devices, to investigate how the electric and magnetic fields affect the electronic properties of double conical surfaces. We expect to explore this problem in forthcoming publications.

{\bf Acknowledgements}: This work was partially supported by Conselho Nacional de Desenvolvimento Cient\'{i}fico e Tecnol\'{o}gico (CNPq), CAPES and FACEPE.

%
           \bibliographystyle{ieeetr}
            \bibliography{ref}
%
%
%

\end{document}